\documentclass{aa}
\usepackage{natbib}

\def\SB9{$S\!_{B^9}$}
\def\Nsys{2\,386}
\def\Norb{2\,694}
\def\NHIP{1\,829}

\def\NGCVS{737}
\def\MissingPapers{537}
\def\AddedPapers{372}
\bibpunct{(}{)}{;}{a}{}{,}

\usepackage[dvips]{graphicx}
\usepackage{times}
\newcommand\ignore[1]{} 

\begin{document}
\title{{\SB9}: The Ninth Catalogue of Spectroscopic Binary Orbits}
\author{D.~Pourbaix\inst{1,2}\fnmsep\thanks{Research Associate, F.N.R.S., Belgium}
\and A.A.~Tokovinin\inst{3}
\and A.H.~Batten\inst{4} 
\and F.C.~Fekel\inst{5}
\and W.I.~Hartkopf\inst{6} 
\and H.~Levato\inst{7}
\and N.I.~Morrell\inst{8}\fnmsep\thanks{on leave from Facultad de Ciencias Astronomicas y Geofisicas, Universidad  Nacional de La Plata; Member of CONICET, Argentina}
\and G.~Torres\inst{9}
\and S.~Udry\inst{10}}
\institute{Institut d'Astronomie et d'Astrophysique, Universit\'e
Libre de Bruxelles, CP. 226, Boulevard du Triomphe, B-1050
Bruxelles, Belgium
\and
Department of Astrophysical Sciences, Princeton University,
Princeton, NJ 08544-1001, USA
\and
Cerro Tololo Inter-American Observatory, Casilla 603    La Serena, Chile
\and
Herzberg Institute of Astrophysics, Dominion Astrophysical Observatory, 5071 West Saanich Road, Victoria, B.C., V8X 4M6, Canada
Victoria
Inst4
\and
Center of Excellence in Information Systems, Tennessee State University, 330 10th Avenue North, Nashville, TN 37203-3401, USA
\and
US Naval Observatory, 3450 Massachusetts Avenue, NW, Washington DC 20392-5420, USA
\and
Complejo Astronomico El Leoncito (CASLEO), Casilla de Correo 467, Av. Espa\~{n}a 1512 Sur, 5400 San Juan, Argentina.
\and
Las Campanas Observatory, The Carnegie Observatories, Casilla 601,
La Serena, Chile.
\and
Harvard-Smithsonian Center for Astrophysics, 60 Garden St., Cambridge, MA 02138, USA
\and
Observatoire de Gen\`eve,Chemin des Maillettes, CH-1290 Sauvergny, Switzerland
}

\date{Received date; accepted date}

\abstract{The Ninth Catalogue of Spectroscopic Binary Orbits
(http://sb9.astro.ulb.ac.be) continues the series of compilations of
spectroscopic orbits carried out over the past 35 years by Batten and
collaborators.  As of 2004 May 1st, the new Catalogue holds orbits
for \Nsys\ systems.  Some essential differences between this catalogue
and its predecessors are outlined and three straightforward
applications are presented: (1) Completeness assessment: period
distribution of SB1s and SB2s; (2) Shortest periods across the H-R
diagram; (3) Period-eccentricity relation.}

\maketitle

\section{Introduction}

Over the past fifteen years the Eighth Catalogue of the Orbital
Elements of Spectroscopic Binary Systems \citep[SB8,
][]{Batten-1989:a} has been used extensively by the community for a
variety of purposes ranging from binary statistics to target
selection.  It is referenced by more than one hundred papers in the
literature.  The progress of the spectro-velocimeters (CORAVEL, CfA
speedometers, etc.), which have significantly increased the number of
known late-type spectroscopic binaries, combined with the need for
refined statistics, have made the revision of the 8th catalogue worth
undertaking.  At the 2000 IAU General Assembly in Manchester,
Commission 30 decided to take responsibility for the 9th catalogue
(\SB9).  We report on the progress achieved so far. The main goal is
to make potential users aware of this new database and its features.
In the process of compiling it, we have also found that it often
serves to correct mistakes of various kinds that have gone unnoticed
in the original publications. 

Catalogues of spectroscopic binary orbits are used in many very
different ways.  For example, they serve to select specific types of
binaries for further observations, such as the compilation by
\citet{Taylor-2003:a}, which is aimed at identifying interferometrically
resolvable systems.  A combination of SB8 with visual-binary and other
catalogues has led to the creation of the database on
high-multiplicity stars \citep{Tokovinin-1997:a}.  Spectroscopic
binaries with detectable motion received special processing during
Hipparcos data reduction \citep{Hipparcos}, and this is anticipated to
be even more important for the next generation of astrometric
satellites such as Gaia \citep{Pourbaix-2003:c} or SIM \citep{Marr-2003:a}.  
A catalogue also enables various statistical studies to be made.  In the 
past, controversial results on the mass ratio distribution were obtained 
from SB catalogues \citep[see the discussion in][]{Duquennoy-1991:b}, and in
recent years volume-limited samples have been much preferred for this
kind of analysis \citep[e.g., ][]{Heacox-1998:a,Halbwachs-2003:a}.
Despite strong and poorly known selection effects, SB catalogues are
nevertheless indispensable for certain types of statistical work where
the large number of objects is of primary importance. This is the
case, for example, when there is a need to define various boundaries
in the orbital-parameter space or when only a small sub-sample of
objects with specific properties is studied, such as chemically
peculiar stars or late-type giants \citep{Boffin-1993:a}.  Finally,
binaries with unusual properties are best found in such large
catalogues. 

The content of \SB9\ in terms of orbits, and the matching of \SB9\
with other major catalogues is described in Sect.~\ref{sect:content},
while the differences with respect to previous compilations are
outlined in Sect.~\ref{sect:diff}.  Sect.~\ref{sect:access} presents
two different ways of accessing the data.  We conclude in
Sect.~\ref{sect:applications} with three applications, two of which
rely upon \SB9\ only, while the other illustrates the benefit of joining
it with other major catalogues such as Hipparcos. 
 
\section{Content}\label{sect:content}

\begin{figure*}[htb]
\resizebox{0.49\hsize}{!}{\includegraphics{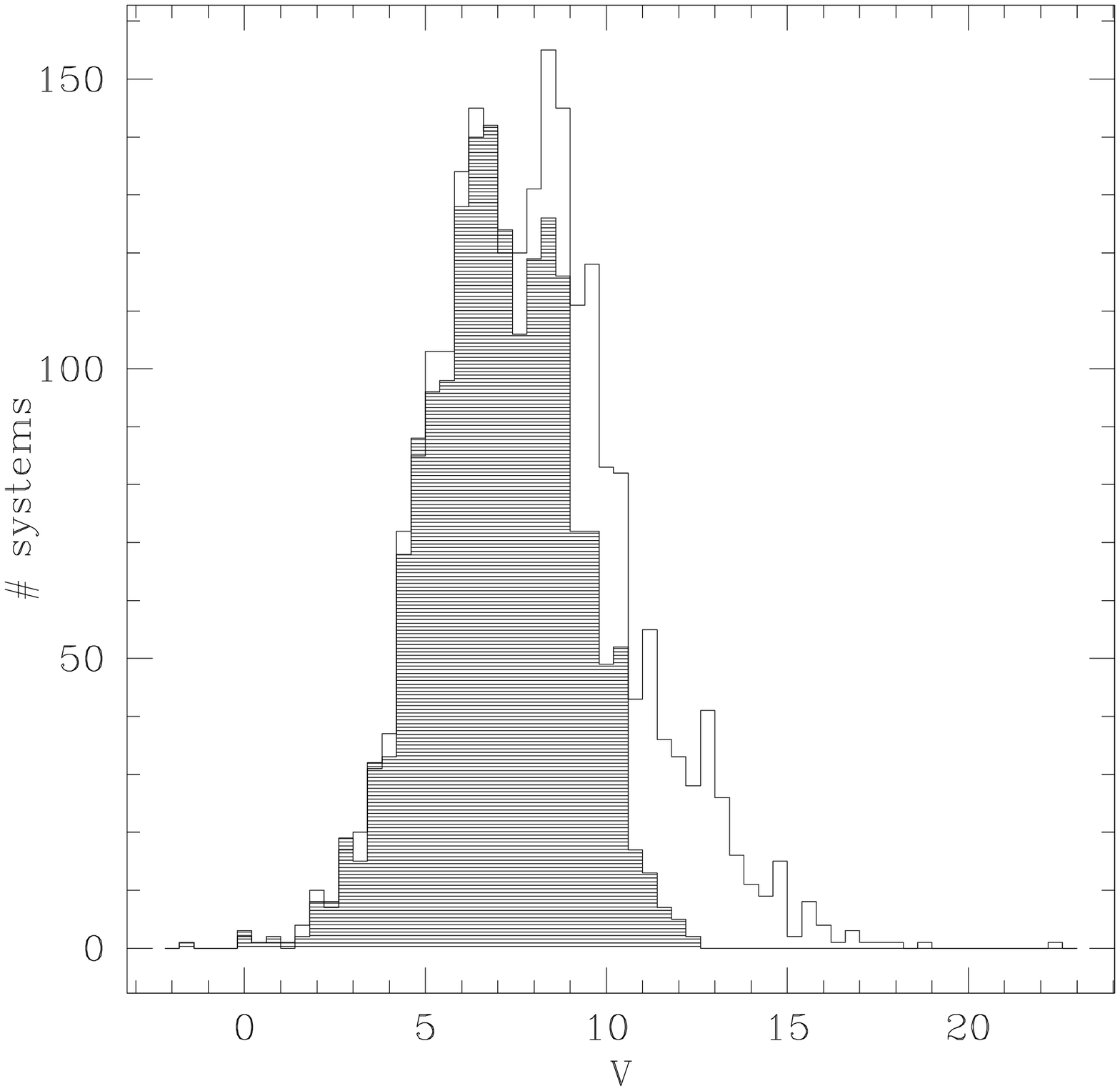}}
\hfill
\resizebox{0.49\hsize}{!}{\includegraphics{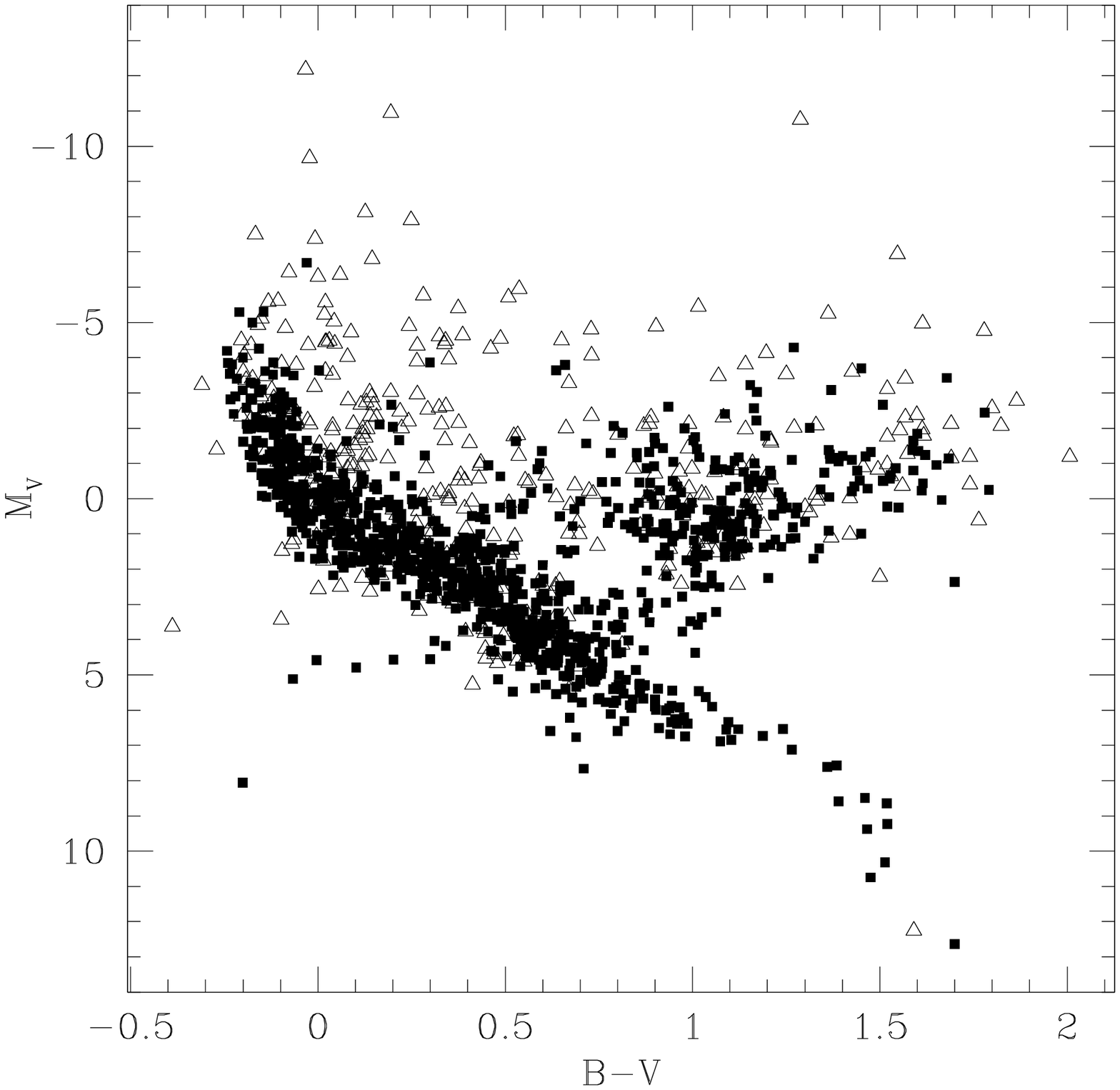}}
\caption[]{\label{Fig:Mag}{\bf Left panel:} Distribution of the
apparent $V$ magnitude for all \SB9\ systems.  
The shaded histogram represents Hipparcos systems.
{\bf Right panel:} H-R
diagram of the \NHIP\ \SB9\ systems after Hipparcos.  Filled squares
denote $3\sigma$ parallaxes whereas open triangles denote less
reliable ones.}
\end{figure*}

By definition the content of \SB9\ changes almost continuously, and
users always access the latest version in real time.  Let us
nevertheless freeze the content at, say, May 1st 2004, i.e., almost
three years after SB8 was used as a seed for its successor. 

\SB9\ contains \Nsys\ systems (versus 1\,469 in SB8) for a total of
\Norb\ orbits (Sect.~\ref{sect:diff} explains the reason of the difference between the number of systems and the number of orbits).
A system can be represented by either two stars or one
star and the center of mass of another system.  A triple star is thus
seen as two distinct systems identified through the component
descriptor.  There are 55 spectroscopic triple (or higher
multiplicity) systems and eight systems with four orbits. 

In terms of cross-references with other catalogues, \NGCVS\ \SB9\
systems are included in the GCVS \citep{Kholopov-1998:a} and \NHIP\ in
the Hipparcos catalogue \citep{Hipparcos}.  The large overlap with
Hipparcos is a selection effect, as illustrated in the left panel of
Fig.~\ref{Fig:Mag}; so far, \SB9\ systems are predominantly bright
objects.  By incorporating the Hipparcos parallaxes, the spread of the
systems over the Hertzsprung-Russell (H-R) diagram is shown in the
right panel of Fig.~\ref{Fig:Mag}. 

The completeness of the database can be evaluated through the number
of published orbits that are still missing.  According to the
Bibliographic Catalogue of Stellar Radial Velocities
\citep{Malaroda-2003:a}, about \MissingPapers\ papers with orbital
solutions published prior to 2003 still need to be uploaded (this is
in addition to the \AddedPapers\ papers already added after SB8).
These \MissingPapers\ papers are expected to yield some 1\,500 orbits
out of which $\sim$1\,200 might be new systems.  While the work of
Malaroda and collaborators is certainly very useful, the actual number
of missing papers is somewhat difficult to evaluate. For example, we
have found and reported to H. Levato a few entries in their catalogue 
that have no orbit in the original publication despite the `ORB' flags
in their entries.  In addition, there is a gap between the date of the most
recent SB8 orbit and the earliest paper in their catalogue.
Sporadically, we still find systems with orbits published much
earlier than the completion of SB8 but which were nevertheless absent
from that compilation. 

\section{Major differences with respect to SB8 and
predecessors}\label{sect:diff}

\begin{figure}[htb]
\resizebox{\hsize}{!}{\includegraphics{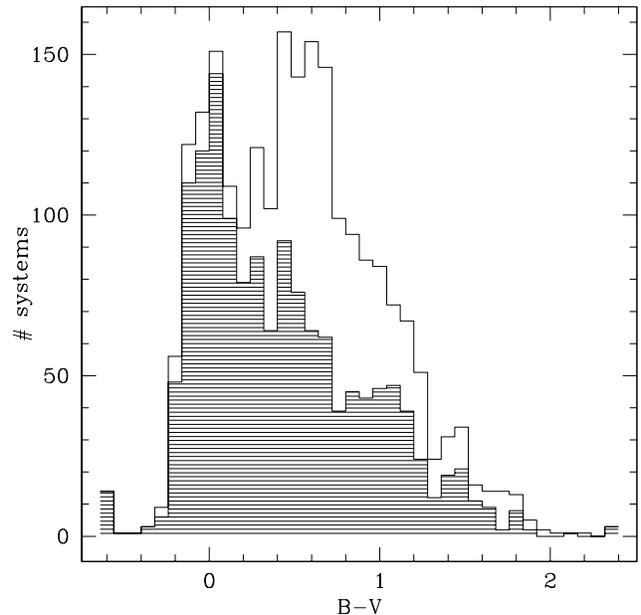}}
\caption[]{\label{Fig:histoBV}Distribution of $B-V$ colors for \SB9\
(resp. SB8) systems as empty (resp. shaded) histograms.}
\end{figure}

For both the SB8 catalogue and the current version of the \SB9\
catalogue we have plotted in Fig.~\ref{Fig:histoBV} the number of
systems versus $B-V$ color.  The primary source for the colors was the
Hipparcos catalogue.  For stars not in that catalogue, the SIMBAD
database was used.  The shaded histogram for SB8 shows 98\% of its
systems.  The completeness for \SB9\ is somewhat less, 94\%, at least
in part because of a significant increase in X-ray binaries.  Despite
this modest disparity in completeness, the comparison of the two
distributions is instructive.  The histogram for the SB8 catalogue
shows a peak in the $B-V$ distribution corresponding to late-B and
early-A type stars, and then a relatively uniform decline in the
number of systems with increasing $B-V$ color.  

The distribution of systems in \SB9\ is rather different.  The strength of 
the peak for the early-type stars is little changed from that in SB8, while
the number of late-type systems has increased dramatically.  The strongest
peak currently occurs for colors that correspond to a range of mid-F
to mid-G stars.  Redward of this peak, there is a significantly
enhanced tail that extends to about $B-V = 1.3$.  This tail contains
numerous evolved stars, including barium stars and chromospherically
active binaries.  At least part of the large increase in the number of
late-type binaries may be attributed to the advent of velocity
spectrometers such as CORAVEL and Griffin's instrument at Cambridge plus
the CfA speedometers, which are most
productive for late-type stars.  With the eventual addition to \SB9\
of several hundred systems that currently have published orbits, we
expect the number of late-type binaries to continue its rapid rise.
Over the next few years orbits for a substantial number of M giant
symbiotic binaries also will be determined, extending the tail. 

In previous versions of the Catalogue the aim was to provide a single
`best' orbit per system even though several different orbits were
often discussed in the notes. \SB9\ plans to list all available orbits,
whether they turn out to be wrong, preliminary or definitive.  The
grade associated with a given orbit (following the practice in
previous catalogues) should help the user select the best
solution if only one orbit is needed.  While the grade in the previous
catalogues did rely upon the expertise of their authors only, we plan to 
derive a much more objective grading scheme (as achieved by the 
US Naval Observatory people for the visual orbits \citep{Mason-2001:c}).  
The improved grading system has not yet been implemented, and so 
currently, the user can form his/her own opinion as to the quality
of a new orbit, based on the uncertainties of the elements and the 
plot of its velocities.

The number of identifiers that can be stored for a given system in \SB9\ 
is unlimited, unlike previous editions, although the major catalogues 
(e.g., HD, DM, HIP, \dots) are still favored as the source for stellar 
identification. 

For the present Catalogue the 2000.0 epoch/equinox was adopted rather
than 1900.0, and the number of digits in the coordinates was also
increased to match the precision adopted by other
compilations \citep[e.g., the Washington Double Star
Catalog;][]{Mason-2001:c}.  Positions are now given to the hundredth of
a second of time in Right Ascension and to the tenth of a second of
arc in Declination.  The 1900.0 coordinates initially kept for
backward compatibility have now been discontinued. 

The orbital parameters are listed in the Catalogue together with their
published uncertainties. For some double-lined systems where separate
values of the eccentricity and the systemic velocity were reported for
each component in the original publication, it was decided to store
only one set, even though this may cause difficulties in some cases.
The second set of values is usually listed in the notes. 

The main difference between \SB9\ and its predecessors has to do with
the data used to derive the orbits.  Whenever possible, all the radial
velocities used for an orbit are also stored in \SB9.  Situations
where this is not possible are already foreseen (e.g., orbits fitted
directly to spectra, without the determination of radial velocities,
etc.).  Nevertheless, an effort is made to be as complete as possible
for orbits published until now, and radial velocities for 1\,113
systems have been included even for some already present in SB8.
Although collecting individual radial velocities is very time
consuming, their availability makes it possible to double-check both
the transcribed orbit and the one published in the literature.
When possible, authors are notified when typographical errors are 
noticed in the original paper.

Since more than one orbit can be listed, it is possible to identify
which systems from SB8 have been reinvestigated.  A total of 177 SB8
systems are listed with two or more orbits, and 13 of them had orbits
which were already considered definitive in SB8 (55 entries 
had such a top grade in SB8). Thus,
$\sim$25\% of the best systems have been investigated again.
Unfortunately, at the other end of the quality range only 6\% of the
low quality orbits have been revised.  One might therefore conclude
that users are more interested in identifying new systems than in
refining those already known but still poorly modeled. 

The bibliographic reference for an orbit is given by the 19-character
bibliography code ("bibcode") used by the NASA Astrophysics Data
System (ADS).  When no bibcode is available for an orbit, a special
code is substituted to indicate that the reference is listed in the
Notes part of the Catalogue. 

\section{Access}\label{sect:access}

A direct consequence of the unlimited number of orbits and identifiers
that can be stored was the replacement of the single table used in SB8
with a relational database structure (several tables).  Whereas a
basic selection (or simply printing) is perhaps more difficult than
with a unique table, such a structure is much more flexible. 

The most convenient way to access \SB9\ and to browse a specific orbit
is through its web interface\footnote{http://sb9.astro.ulb.ac.be/}.
Systems can be searched by catalogue identifier as well as by
coordinates.  If several systems match the selection criteria, the
user is asked to further select one.  If several orbits are available,
the user selects one of them for display.  The year of publication is
listed to aid in making a choice among the orbits. 

The displayed information takes advantage of HTML by offering links.
In addition to the coordinates, spectral type, apparent magnitude,
identifiers, orbital parameters, and other information, the interface
offers a direct link to ADS thus allowing straightforward retrieval of
the abstract of the paper.  The interface also offers the automatic
plotting of the orbit (with the actual observations, if available).
The corresponding figure is displayed on the screen, and a PostScript
version is available as a link. 

For researchers interested in properties of a sample of these systems
rather than in browsing one orbit, a compressed `tar ball' version of
part of the database is also available from the \SB9\ main page.  Only
the radial velocity files are missing from that archive.  Coupled with
Unix-like standard tools such as {\em sort} and {\em join}, and popular 
scripting languages such as {\em awk} and {\em python}, the possibilities
offered by this distribution of the database combined with other public access
catalogues are almost unlimited, as illustrated in
Sect.~\ref{sect:applications}. 

\section{Applications}\label{sect:applications}

\subsection{Completeness assessment}

Although as mentioned in Sect.~\ref{sect:content} completeness of
\SB9\ has not yet been achieved, it is still possible to evaluate its
statistical completeness. If one aims at a database useful for
statistical purposes, the extent to which the present orbits represent
the parent population is what matters. 

\begin{figure}
\resizebox{\hsize}{!}{\includegraphics{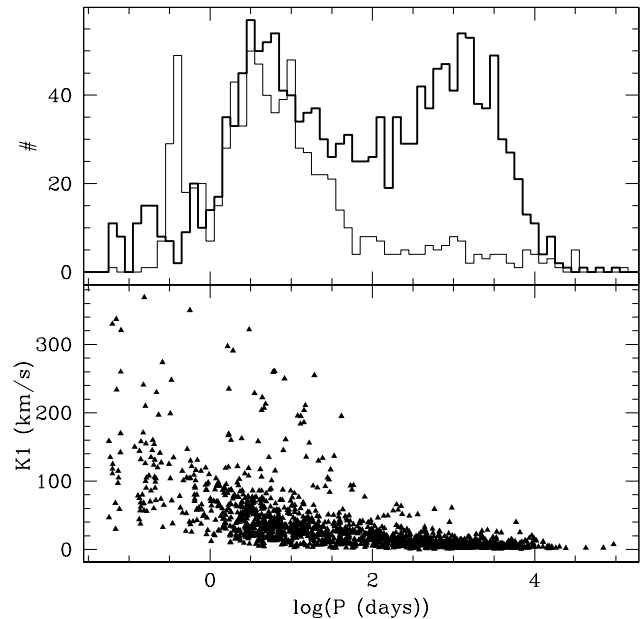}}
\caption[]{\label{Fig:distP}{\bf Top panel:} Distribution of the
orbital periods in days.  The thick (thin) line histogram represents
the period of single-lined (double-lined) systems. {\bf Bottom panel:}
Amplitude of the radial velocity curve as a function of the orbital
period for single-lined systems.}
\end{figure}

In their statistical studies of F7-K binaries with periods under 10
years, \citet{Halbwachs-2003:a} obtained a bi-modal distribution of
the orbital period with peaks at 20 days and $\sim$2 years.  Owing to
the size of their sample, they could not rule out the possibility that
the distribution was actually consistent with a log-normal
distribution.  In the top panel of Fig.~\ref{Fig:distP} we show the
distribution of the periods for both the single-lined (SB1) and
double-lined (SB2) systems in \SB9\ (the latter accounting for 1/3 of
the entire database).  For single-lined systems the distribution is
clearly bi-modal, with peaks at $\sim$4 days and $\sim$1\,200 days.
While the distribution of periods for SB2 systems appears bi-modal as
well, the two only share one peak, since the SB2 distribution
has a peak at $\sim$0.4 days corresponding to contact binaries. 

Can the absence of a peak at $\sim$1200 days in the distribution of
SB2 periods be explained as an observational bias?  Long orbital
periods imply small velocity amplitudes, which are more difficult to
detect. To illustrate this, the lower panel of Fig.~\ref{Fig:distP}
shows the distribution of semi-amplitudes $K_1$ for SB1 systems. It is
seen that even for periods in excess of 1000 days the semi-amplitudes
remain large compared to typical measurement errors in the velocities,
so in principle such orbits should still be detectable for SB2s.
However, the main difficulty for long-period double-lined systems is
blending of the spectral lines.  The low amplitudes make it much more
difficult to detect the lines of the two components in the first
place, let alone disentangle them and measure their velocities. This
observational effect is likely to contribute to the paucity of SB2s
with long periods.  The lack of a peak at 0.4 days for SB1s, however,
seems more difficult to explain as an observational bias. 

Another striking feature of the bottom panel of Fig.~\ref{Fig:distP}
is the behavior of $K_1$ for periods below 3 days ($\log P<0.5$).
Whereas the distinction between SB1s and SB2s is often just a matter
of the magnitude difference between the components, the value of $K_1$
for a given SB1 system, $e$ and $P$ fixed, is a proportional to $\sin i$, $i$ being the
orbital inclination. Any value of $K_1$ down to zero is equally likely 
because a random orientation of orbital planes corresponds to a uniform 
distribution of $\sin i$ in the $[0,1]$ interval. 
There is a rather clear lack of SB1 systems in the lower left corner of the
diagram, although it is too early to tell whether this is real, or the 
result of observational bias. 

\subsection{H-R diagram and the shortest periods}

While \SB9\ contains only limited information on each system aside
from the orbital parameters (whereas other catalogues such as that by
\citet{Taylor-2003:a} list many more properties), its plain text
format with simple field delimiters makes it straightforward to join
with other tables, e.g., the Hipparcos and Tycho Catalogues
\citep{Hipparcos}. As an example, let us consider the location of all
the systems in the color-magnitude diagram based on the Hipparcos
parallaxes and colors, along with the periods from \SB9. What is the
shortest possible orbital period for a binary across the H-R diagram?
Clearly this period depends on the radius of the star (Roche lobe
filling).  \SB9\ makes it possible to investigate this interesting
question. 

\begin{figure} \resizebox{\hsize}{!}{\includegraphics{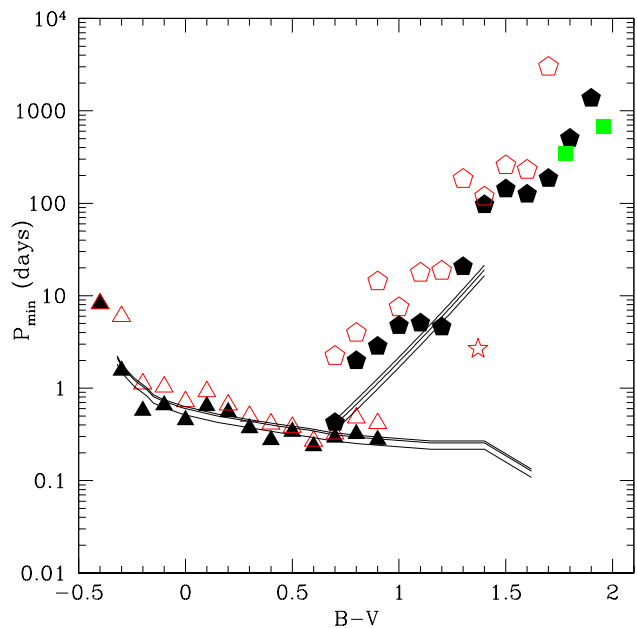}}
\caption[]{\label{Fig:Pmin}Minimum and third shortest periods (filled
and open symbols, respectively) as a function of $B-V$.  Pentagons are
for systems not belonging to the main-sequence according to their
$B-V$ ($>0.7$) and their absolute magnitude ($M_V<4$).  The two filled
squares are suspected giants missing from the Hipparcos catalogue 
\citep{Hipparcos}.  The open pentagon/star is the triple system HIP~9640.} 
\end{figure}

In Fig.~\ref{Fig:Pmin} we display the shortest periods for each 0.1-mag
bin of $B-V$.  Filled symbols represent the minimum period while the open
symbol stands for the third shortest period in that bin.  The distance
between the two symbols therefore gives an estimate of the confidence in
that minimum period.  For $B-V>0.7$ there is a distinction between
main sequence and giant stars based solely on the absolute magnitude:
$M_V>4$ is considered main-sequence (we make no distinction here
between giants and super-giants).  Triangles/pentagons represent
main-sequence/giant stars.  Owing to the lack of main-sequence stars
redward of $B-V=1$, we omit those points.  Two stars thought to be
giants but absent from the Hipparcos catalogue are plotted as filled
squares. 

The tracks in the figure represent the theoretical periods
corresponding to systems where one component fills its Roche lobe,
with a secondary mass of $0.2m_1$, $0.6m_1$, and $0.9m_1$, where $m_1$
is the mass of the primary.  For the main-sequence curves we use the
data after \citet{Popper-1980} together with relations from
\citet{Schmidt-Kaler-1982}.  The giant tracks are based on a rough
estimate of $M_{H_p}$ from Fig.\ 3.5.5 of the Hipparcos catalogue,
with the assumption that $H_p=V$ and a quadratic fit of the bolometric
correction vs.\ $B-V$ from Schmidt-Kaler. 

The agreement between the theoretical tracks and the observations is
excellent with the exception of one giant star (HIP~9640) with $P\sim
3$ days instead of 20 days.  However, this object is a known multiple
system.  Whereas the primary is indeed a supergiant (thus explaining
the pentagon/star symbol), that star does not belong to the spectroscopic
system for which the period is given in \SB9.  Therefore, nothing
prevents the period from being shorter than inferred from the
luminosity class of the whole system. 

\subsection{Period-Eccentricity}

\begin{figure}
\resizebox{\hsize}{!}{\includegraphics{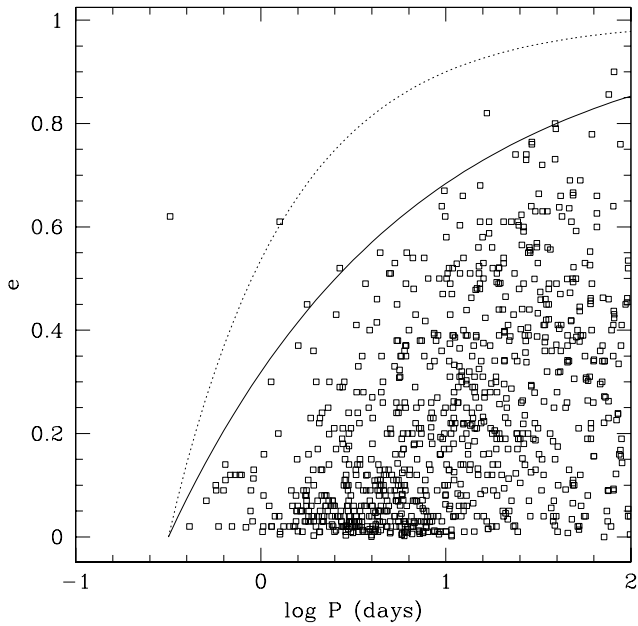}}
\caption[]{\label{Fig:spe}Period-eccentricity relation for all \SB9
objects with $P<100$d and $e>0$ (see text).}
\end{figure}

In Fig.~\ref{Fig:spe} we show the period-eccentricity relation for all
\SB9\ objects with $P<100$ days and non-zero eccentricity. The upper
envelope of the data distribution -- largest eccentricity for a given
period -- is not well understood.  According to
current theory the eccentricity is limited either by contact of the
components or by tidal effects, both being determined by the distance
at periastron $a(1-e) \propto P^{2/3}(1-e)$.  In this case the upper
envelope would be described by a line $P(1-e)^{3/2} = {\rm const}$.
Such a law (dotted line in Fig.~\ref{Fig:spe}) is not a good match to
the data, and the solid line $P(1-e)^{3}={\rm const}$ describes the
envelope much better. The points above the full line correspond to a
pulsar PSR~1913+16 (No.\ 1137 in SB8) and to two orbits of poor quality
that are likely erroneous (No.\ 869 and 353 in SB8).  The upper
envelope is formed by spectroscopic binaries with early-type (O, B, A)
main sequence components, since low-mass binaries tend to be
circularized by tidal forces. 

There is no apparent explanation for the upper envelope. The constraints 
on eccentricity derived by \citet{Hut-1981:a} from dissipative tidal evolution 
lead to a lower limit on the angular momentum $h$. Given that $h^2 
\propto a (1 - e^2)$, we get the upper envelope $P(1-e^2)^{3/2} = {\rm const}$ 
which is similar to the dotted, rather than solid line. Quantities 
such as angular velocity at periastron or the fraction of the orbit 
spent near periastron again correspond to the dotted line.

\section{Conclusions}

Due to our limited manpower for this work and the long delay since the
release of SB8, \SB9\ is still far from achieving an adequate degree
of completeness.  For instance, no orbit of any extra-solar planet is
present yet.  Despite its `work in progress' status, the present
version of \SB9\ already offers several advantages over SB8, among
which is an increase of 62\% in the number of systems listed. Numerous
applications should benefit from these increased numbers (e.g., more
than 800 double-lined systems await a determination of their
individual masses by interferometric means). 

Authors of orbital solutions are invited (and indeed urged) to send DP
the \TeX\ or \LaTeX\ version of their papers for quick upload into
\SB9.  Tools are also available for those interested in formating
their own data prior to inclusion in \SB9. 

\begin{acknowledgements}

We are pleased to acknowledge contributions from J.-M.~Carquillat, 
E.V.~Glushkova, R.F.~Griffin, M.~Imbert, A.~Jorissen, R.~Leiton, 
D.~Stickland, L.~Szabados, and J.~Tomkin.  This work is partly
supported by NASA grant NAG5-11094 to Princeton University.  Additional
support has been provided by NASA grant NCC5-511 and NSF grant HRD-9706268
to Tennessee State University.  This Catalogue has made use of the SIMBAD 
database, operated at CDS, Strasbourg, France. 

\end{acknowledgements}

\bibliographystyle{aa} 
\bibliography{articles,books}

\end{document}